\documentclass[%
mathleft,%
]{an}
\usepackage{graphicx}
\usepackage{times}
\begin{document}

% The following seven commands are intended for editorial usage and
% should be ignored by the author(s).
\Pagespan{1}{}% Document's page range. 
% If second parameter is left empty, the last page is computed
% automatically.
\Yearpublication{2011}%
\Yearsubmission{2011}%
\Month{1}%   
\Volume{999}%  
\Issue{92}% 
% \DOI{This.is/not.aDOI}% 

\title{﻿Dichotomy in the population of young AGN: optical, radio and X-ray properties}

\author{Magdalena Kunert-Bajraszewska\inst{1}\fnmsep\thanks{Corresponding author:
  \email{magda@astro.uni.torun.pl}}
% Example for footnote, note the usage of the \texttt{fnmsep} command
% as separator between institute number and footnote mark}
%\and  G.\,H. Ostwriter\inst{2,3}
}
\titlerunning{Dichotomy in the population of young AGN}
\authorrunning{Magdalena Kunert-Bajraszewska}
\institute{
Toru\'n Centre for Astronomy, Faculty of Physics, Astronomy and Informatics, NCU, Grudziadzka 5 , 87-100 Toru\'n, Poland}

\received{XXXX}
\accepted{XXXX}
\publonline{XXXX}


\abstract{There are numerous examples of radio sources with various sizes which surprisingly exhibit very 
similar morphology. This observational fact helped to create a standard evolutionary model in which young and small radio-loud active galactic nuclei (AGN) called gigahertz-peaked spectrum (GPS) sources and compact steep spectrum (CSS) sources, become large-scale radio objects. However, many details of this evolutionary process are still unclear.  We explored evolution scenarios of radio-loud AGN using new radio, optical and X-ray data of so far unstudied low luminosity compact (LLC) sources and we summarize the results in this paper. Our studies show that the evolutionary track is very 'personalized' although we can mention common factors affecting it. These are interaction with the ambient medium and AGN power. The second feature affects the production of the radio jets which if they are weak are more vulnerable for instabilities and disruption. Thus not all GPS and CSS sources will be able to develop large scale morphologies. Many will fade away being middle-aged ($10^5$ years). It seems that only radio strong, high excitation compact AGN can be progenitors of large-scale FR\,II radio sources.}

\maketitle

\section{Introduction}
Evolutionary paths of radio sources are commonly plotted in a radio power - linear size plane. This parametrization of the radio source behaviour does serve to capture the major evolutionary trends that can be compared with observations since it involves the two most evident, easily measured features of extended radio sources. In the general scenario of the evolution of powerful radio-loud AGN (Readhead et al. 1996; Fanti et al. 1995; O'Dea \& Baum 1997), the younger and smaller gigahertz-peaked spectrum (GPS) and compact steep spectrum (CSS) sources become large scale FR\,I and FR\,II objects (Fanaroff \& Riley,  1974). 

However, the growing number of observations of low power radio sources and results of their analysis suggest that not one but at least two evolutionary paths exist (Snellen et al. 2000; Ma\-rec\-ki et al. 2003; Ku\-nert-\-Baj\-ra\-szew\-ska et al. 2010; An \& Baan 2012; Maciel \& Aleksander 2014; Turner \& Shabala 2015). The determining factors for the evolution of radio objects can be related either to jet-interstellar medium (ISM) interactions or central engine (differences in the accretion mode or black hole spin, instabilities in the accretion flow).  
During their evolution the radio jets start to cross the ISM and try to leave the host galaxy. The interaction with the ISM can be very strong in GPS and CSS sources and it seems to be a
crucial point in the evolution of radio sources (Labiano 2008; Holt, Tadhunter \& Morganti 2009; Kawakatu et al. 2009; Perucho, Quilis \& Mart\'i 2011; Wagner et al. 2012; Chandola, Gupta \& Saikia 2013; Dallacasa et al. 2013; Maccagni et al. 2014; Ger\'eb et al. 2015). 

On the other hand the study of the optical properties of radio-loud AGN suggests the existence of two radio source populations where the main discriminant is the accretion rate onto the central black hole (Hardcastle, Evans \& Croston 2007; Buttiglione et al. 2010; Ku\-nert-Baj\-ra\-szew\-ska \& Labiano 2010; Best \& Heckman 2012).  According to these studies high excitation galaxies (HEG) are distinctive in strong evolution while the low excitation galaxies (LEG) are a slowly-evolving or non-evolving group of radio-sources. This has also been recognised by other authors who suggest that some young radio-loud AGN can be short-lived objects on time scales $10^4 - 10^5$ years (Reynolds \& Begelman 1997; Aleksander 2000; Ma\-rec\-ki et al. 2003; Czerny et al. 2009; Ku\-nert-\-Baj\-ra\-szew\-ska et al. 2010, Shulevski et al. 2015). Very recently, based on the optical and X-ray observations, Scha\-winski et al.(2015) concluded that the whole lifetime of AGN consists of numerous short phases (each phase typically lasts for
$\sim 10^5$ years) alternating between high and low accretion rate.

The analysis of the radio, optical and X-ray properties of a sample of low luminosity compact (LLC) sources carried out by us over the last few years provides new information in this
active field of research. These studies
brought new interesting clues on the evolution of radio-loud AGN (Ku\-nert-Baj\-ra\-szew\-ska \& Labiano 2010; Ku\-nert-\-Baj\-ra\-szew\-ska et al. 2010, 2014) and we shortly summarize them here.

\section{Our sample}
\begin{figure}
\includegraphics[width=\columnwidth]{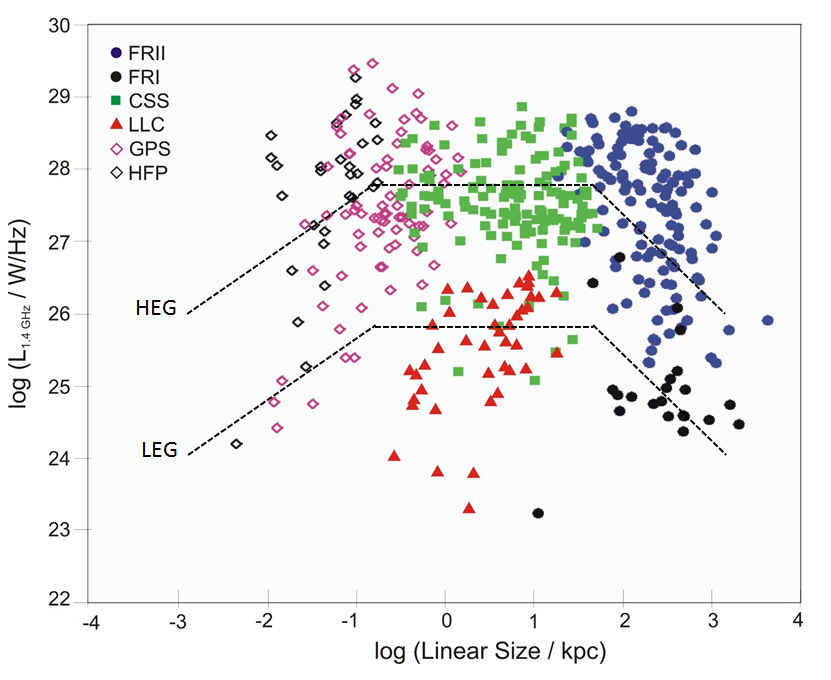}
\caption{Radio power -- linear size diagram for radio-loud AGN. The new sample of low luminosity compact (LLC) sources are indicated with red triangles. High frequency peakers (HFP) are a sub-population of compact AGN defined by Dallacasa et al.(2000). Additionally two independent evolutionary tracks for HEG and LEG sources are outlined (see section \ref{accretion}).}
\label{radio}
\end{figure}

The main aim of our project was to explore the lower luminosity region of the radio power - linear size plane occupied by compact radio sources. Thus, the most important criterion for selecting the sample was the radio luminosity threshold.
To look for these objects we combined information from the {\it Faint Images of the Radio Sky at 20-cm (FIRST)} (White et~al. 1997), the {\it Green Bank 6-cm (GB6) survey} (Condon et al. 1994) 
and {\it Sloan Digital Sky Survey (SDSS)}.  
The selection criteria we used (for details see Ku\-nert-\-Baj\-ra\-szew\-ska et al. 2010) resulted in approximately one-third of the sources having a value of 1.4\,GHz radio luminosity comparable to FR\,Is. The final sample consisted of 44 LLC objects. They are in poorly studied to date area of the radio power - linear size plot (Figure \ref{radio}). The radio observations of LLC objects were carried out with MERLIN at L-band and C-band in a few observational campaigns. 

As the next step we examined the optical properties of LLC sources and applied for them the division for low and high excitation galaxies (Ku\-nert-Baj\-ra\-szew\-ska \& Labiano 2010). The obtained results are in agreement with a dual-po\-pu\-lation model and evolution for large scale radio AGN (Buttiglione et al. 2010; Best \& Heckman 2012). 

And finally, the new {\it Chandra} observations of some LLC objects revealed interesting individual cases (Ku\-nert-\-Baj\-ra\-szew\-ska, Siemiginowska \& Labiano 2013) and helped us with further analysis of the properties and evolution of CSS sources (Ku\-nert-\-Baj\-ra\-szew\-ska et al. 2014).

\section{Discussion}
\subsection{Impact of environment ...}

The majority of LLC sources were resolved with multi-ele\-ment radio linked interferometer network (MERLIN) and revealed many types of radio structures.
A large percentage ($\sim 86$ per cent) of objects with a detected core show
difference in the brightness of the lobes located on both its sides. 
The same result has been previously found for the CSS source 
population studied by Saikia et~al.(2001). It is probably due to the fact that a young source is confined within the host galaxy and needs to struggle with the interstellar medium during the expansion (Labiano 2008; Holt, Tadhunter \& Morganti 2009; Kawakatu et al. 2009; Chandola, Gupta \& Saikia 2013; Dallacasa et al. 2013; Ger\'eb et al. 2015). If during the evolution jets are expanding in an inhomogeneous ambient medium as an effect we observe disrupted radio structure which can be preserved to larger scale (Wagner et al. 2012). The most extreme cases of disturbed large scale structures belong to hybrid morphology radio sources (HYMORS). This is a rare class of double-lobed radio sources where each of the two lobes clearly exhibits a different FR morphology (Gopal-Krishna \& Wiita 2000; Gaw\-ro\'n\-ski et al. 2006). Recent high resolution radio observations of HYMORS showed that their parsec-scale structures are similar to that of FR\,II sources. Thus the different radio behaviour that we observe must arise on a larger, probably kiloparsec-scale (Ceg\l{}owski et al. 2013), where the jets of young AGN try to break through the ISM. 

One exception to the above interpretation is the binary quasar FIRST J164311.3+315618 that we found in our sample (Ku\-nert-Baj\-ra\-szew\-ska \& Janiuk 2011). Its complex and distorted radio and optical morphology indicates the intermittent activity with a possible rapid change of the jet direction and/or restarting of the jet due to the interaction with the companion.

\begin{figure}
\includegraphics[width=\columnwidth]{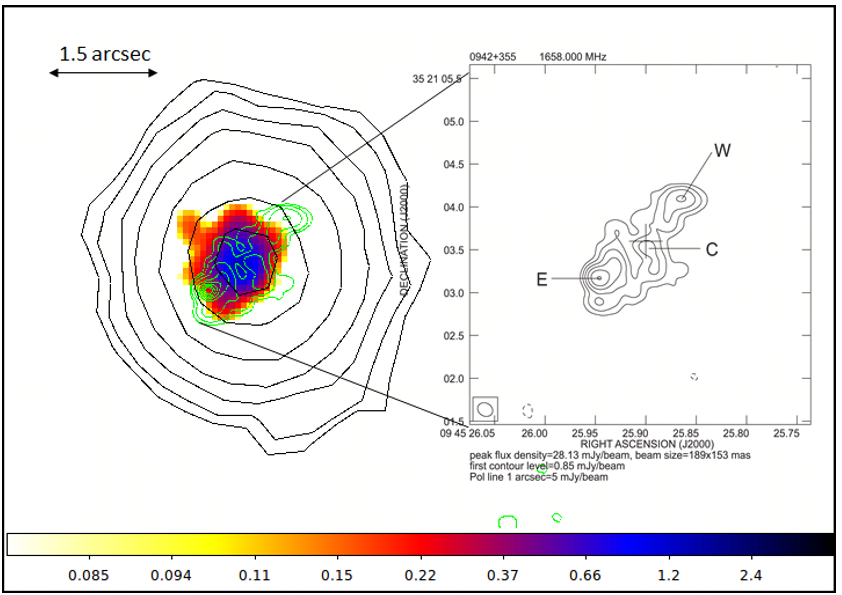}
\caption{SDSS optical (black contours), {\it Chandra} X-ray (color) and MERLIN 1.6\,GHz (green contours) emission from 0942+355. Additionally the enlarged radio map is also shown. We have used 0.15 pixel blocking for the X-ray image.
The cross indicates the position of an optical counterpart taken from the
SDSS. The radio contours increase by factor of 2, the first contour level   
corresponds to $\approx3\sigma$ image noise and amounts to 0.85 mJy/beam. }
\label{all}
\end{figure}

Another distinctive feature of LLC sources is the fact that about 30\% of them have
weak extended radio emission and breaking-up structures. The weak radio core emission is also present and in some cases it can stand for a large fraction of the whole emission of the source. We estimated the core dominance parameter $\rm R=S_{core}/(S_{tot}-S_{core})$ for those LLC objects that have C-band observations and well separated radio core in the 5\,GHz image (0854+210, 0923+079, 1154+435, 1156+470, 1506+345). They have on average ($\rm R_{ave}=0.23$) higher core dominance than larger FR\,IIs ($\rm R_{ave}=0.02$) and FR\,Is ($\rm R_{ave}=0.04$; Morganti, Killeen, Tadhunter 1993) and strong CSS sources ($\rm R_{ave}<0.1$; Saikia et~al.2001). The statistics is low but seems to be in agreement with recently reported high values of R for the more radio weak FR\,0 sources (Ghisellini et al. 2014; Sadler et al. 2014; Baldi, Capetti, Giovannini 2015). These objects similarly to some LLC sources are not able to develop strong extended radio structures for some reason. As we discussed in Ku\-nert-\-Baj\-ra\-szew\-ska et al.(2010) different mechanisms and interactions may play an important role here. This could be the already mentioned jet-ISM interactions but also differences in the properties of the so called central engine: black hole spin or accretion mode. The later can cause the production of weaker jets that are more vulnerable to instabilities and disruption. 
It has already been recognised by some authors (Hardcastle, Evans \& Croston 2007; Buttiglione et al. 2010; Best \& Heckman 2012) that once we take into account the optical spectrum (accretion mode) of the FR\,I and FR\,II sources we can talk about two sub-population of radio-loud AGN: the high and low excitation galaxies (HEG and LEG, respectively). We have extended these optical studies to CSS and GPS sources.

\subsection{... or difference in the accretion mode? }
\label{accretion}

Optical spectroscopic information can play an important role in gaining a better understanding of the properties of central engines of radio sources. The emission-line diagnostic diagrams are able to distinguish H\,II regions ionized by young stars from gas clouds ionized by nuclear activity. Additionally, emission line luminosities show broad connection with radio power what points to a common energy source for both (Buttiglione et al. 2010; Best \& Heckman 2012). 

SDSS optical data are available for most of the LLC sources and using the emission-line ratios we have classified the sources as HEGs and LEGs (Kunert-Bajraszewska \& Labiano 2010).  
We have compared
the [O III] luminosity with the radio properties for LLC sources,  and expanded the sample with
other  CSS,  GPS  sources  and  FR\,I  and  FR\,II  objects.  
The whole sample shows that the HEG sources are brighter than LEG
in the [O III] line by a factor of 10 notwithstanding their linear size that is the evolutionary stage. 
The main evolution scenario (GPS$\to$CSS$\to$FR\,II) for radio-loud AGN was proposed years ago (Readhead et al. 1996; Fanti et al. 1995; O'Dea \& Baum 1997). However, once
the optical division is included, these sources  seem  to evolve in parallel:  $\rm GPS_{LEG}-CSS_{LEG}-FR_{LEG}$ and $\rm GPS_{HEG}-CSS_{HEG}-FR_{HEG}$ (indicated schematically in Figure \ref{radio}). 
Concerning LEG, it is still not clear if $\rm CSS_{LEG}$ would evolve
directly to FR\,I LEG or go through a FR\,II LEG
phase before the FR\,I LEG.

\begin{figure}
\includegraphics[width=\linewidth]{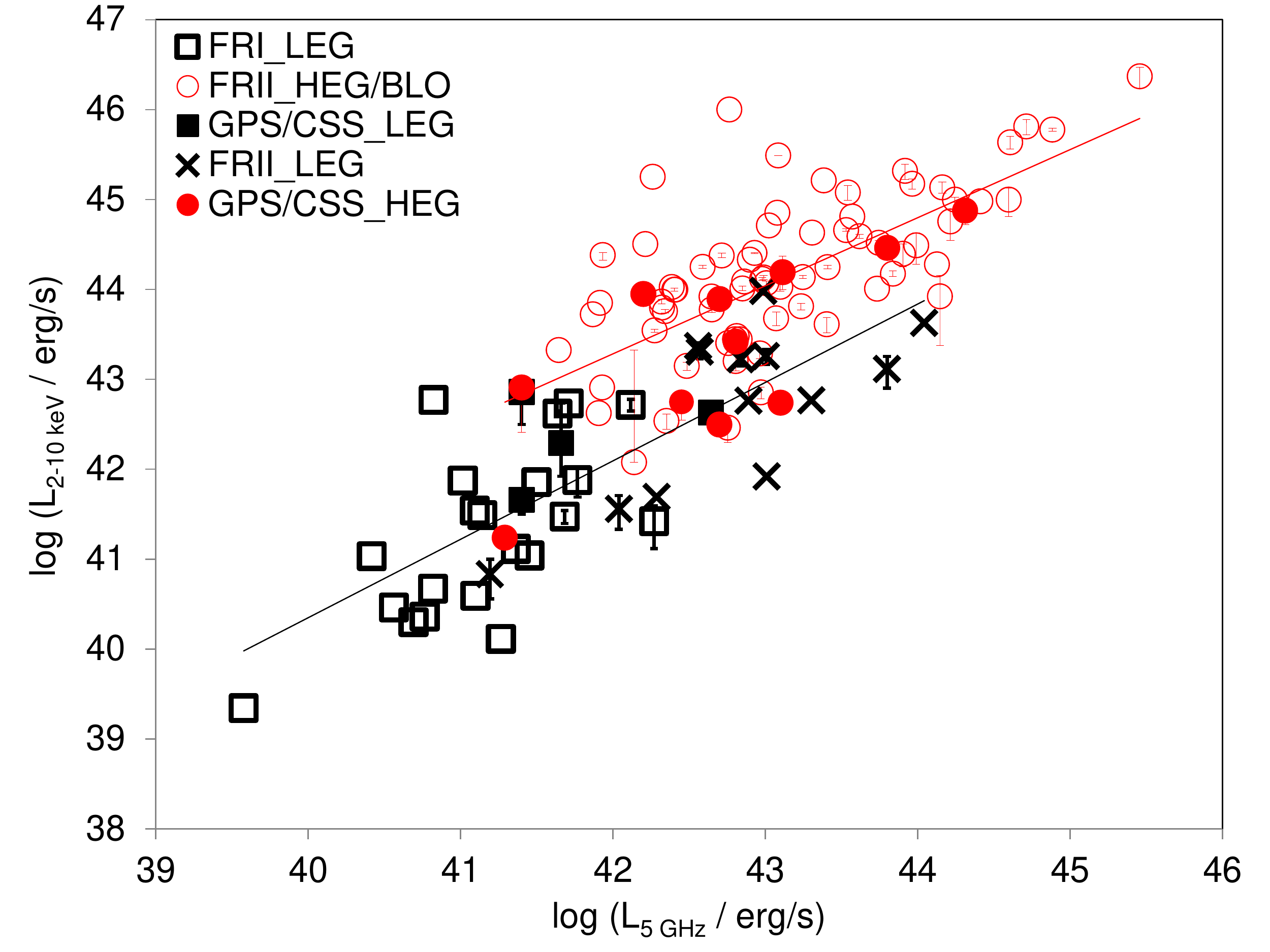}
\caption{2-10\,keV - 5\,GHz luminosity diagram for AGN classified as HEGs and LEGs (reprinted from Kunert-Bajraszewska et al.(2014)). The classification for FR\,I and FR\,II sources were taken from Buttiglione et al. 2010, where the broad line objects (BLO) are considered as members of
the HEG class.}
\label{x-ray}
\end{figure}

\subsection{Radio/X-ray luminosity plane}

A pilot sample consisting of seven LLC sources was selected for X-ray observations using {\it Chandra}.
Four of them have been
detected, the other three have upper limit estimations for X-ray flux (Kunert-Bajraszewska et al. 2014). One of the objects, 1321+045, appeared to be associated with an X-ray cluster (Ku\-nert-\-Baj\-ra\-szew\-ska, Siemiginowska \& Labiano 2013).
The {\it Chandra} ACIS-S image of one of the sources with the largest 
number of X-ray photons is shown in Figure \ref{all} as an example. We also overlayed
the SDSS optical and radio MERLIN 1.6\,GHz contours on the X-ray emission to picture the scale of these emissions. 

We again searched through the literature for other GPS and CSS sources and large scale FR\,Is and FR\,IIs with X-ray detections in order to determine the nature of the relation between morphology, X-ray
properties and excitation modes in radio-loud AGN. We have compared the X-ray luminosity of the radio sources from all above
mentioned groups with their radio properties. 
The large-scale FR\,II sources and strong GPS and CSS objects settle at
higher X-ray and radio luminosities,
while the low power CSSs occupy the space among FR\,I objects that are weaker in X-rays.

Once the HEG/LEG classification is applied also to the X-ray detected AGN, both these groups occupy
distinct locus in the radio/X-ray luminosity plane as shown in Figure \ref{x-ray}.

And finally we have tested the AGN evolution models
by comparing the radio/X-ray luminosity ratio with the size of the sources, and indirectly, with
their age. We found that the less radio powerful FR\,Is have higher radio/X-ray luminosity ratio than many
FR\,IIs what may imply higher X-ray luminosity decrease with radio power in FR\,Is than
FR\,IIs. This is in agreement with previously postulated different X-rays emission origins.
In the case of FR\,Is the X-ray emission is connected with the base of the jet while in FR\,IIs it
originates from accretion (Siemiginowska et al. 2008, Tengstrand et al. 2009). The same can be true for compact sources as well. The results
suggest that below certain level of radio luminosity GPSs and CSSs start to resemble
FR\,Is.

\section{Summary}
The analysis of the radio and X-ray properties of LLC sources together with their spectroscopic features indicate that most of them (especially the HEGs) will evolve finally to FR\,IIs. However, 
they are the tip of the iceberg and much larger population of short-lived low power radio sources are still wating to be explored with high sensitivity radio surveys. They probably belong to the branch of slowly evolving low excitation galaxies and could be the missing precursors of large-scale FR\,I sources.

\acknowledgements

The research leading to these results has received funding from
the  European Commission Seventh Framework Programme (FP/2007-2013)
under grant agreement No 283393 (RadioNet3)

This research has made use of the SDSS. Funding for SDSS-III has been provided by the Alfred P. Sloan Foundation, the Participating Institutions, the National Science Foundation, and the U.S. Department of Energy Office of Science. The SDSS-III web site is http://www.sdss3.org/.

This research has made use of data obtained by the Chandra X-ray
Observatory, and {\it Chandra} X-ray Center (CXC) in the application
packages CIAO, ChIPS, and Sherpa.  This research is funded in part by
NASA contract NAS8-39073. Partial support for this work was provided
by the {\it Chandra} grants GO1-12124X.


\begin{thebibliography}{}
\bibitem{} Aleksander, P.: 2000, MNRAS, 319, 8

\bibitem[An \& Baan(2012)]{an12} An, T., Baan, W. A.: 2012, ApJ, 760, 77

\bibitem{} Baldi, R. D., Capetti, A., Giovannini, G.: 2015, A\&A, 576, 38

\bibitem{} Best, F. N., Heckman, T. M.: 2012, MNRAS, 421, 1569

\bibitem[Buttiglione et al.(2010)]{butti10} Buttiglione S., Capetti, A.,
Celotti, A., Axon, D.J., Chiaberge, M., Macchetto, F.D., Sparks, W.B.: 2010,
A\&A, 509, 6

\bibitem[Ceg\l{}owski et al.(2013)]{ceglowski} Ceg\l{}owski, M.,
Gawro\'nski, M. P., Kunert-Bajraszewska, M.: 2013, A\&A, 557, 75

\bibitem{} Chandola, Y., Gupta, N., Saikia, D. J.: 2013, MNRAS, 429, 2380

\bibitem{} Condon J.J., Broderick J.J., Seielstad G.A., Douglas K., Gregory P.C.: 1994, AJ, 107, 1829

\bibitem[Czerny et al.(2009)]{czerny} Czerny, B., Siemiginowska, A., Janiuk,
A., Nikiel-Wroczy\'nski, B., Stawarz, {\L.}: 2009, ApJ, 698, 840

\bibitem{} Dallacasa, D., Stanghellini, C., Centoza, M., Fanti, R.: 2000, A\&A, 363, 887

\bibitem{} Dallacasa, D., Orienti, M., Fanti, C., Fanti, R., Stanghellini, C.: 2013, MNRAS, 433, 147

\bibitem[Fanaroff \&\-Riley(1974)]{Fanaroff74} Fanaroff, B.~L., \& Riley,
J.~M.: 1974, MNRAS 167, 31

\bibitem[Fanti et al.(1995)]{Fanti95} Fanti, C., Fanti, R., Dallacasa, D., et
al.: 1995, A\&A 302, 317

\bibitem{} Gawro\'nski, M. P., Marecki, A., Kunert-Bajraszewska, M., Kus, A. J.: 2006, A\&A, 447, 63

\bibitem{} Ger\'eb, K., Maccagni, F. M., Morganti, R., Oosterloo, T. A.: 2015, A\&A, 575, 44 

\bibitem{} Ghisellini, G., Tavecchio, F., Maraschi, L., Celotti, A., Sbarrato, T.: 2014, Nature, 515, 376

\bibitem{} Gopal-Krishna; Wiita, P. J.: 2000, A\&A, 363, 507 

\bibitem{} Hardcastle, M. J., Evans, D. A., Croston, J. H.: 2007, MNRAS, 376, 1849

\bibitem{} Holt, J., Tadhunter, C. N., Morganti, R.: 2009, MNRAS, 400, 589

\bibitem[Kawakatu et al.(2009)]{Kawakatu09a} Kawakatu, N, Kino, M., Nagai,
H.: 2009, ApJ, 697, 173L

\bibitem[Ku\-nert-\-Baj\-ra\-szew\-ska \& Thomasson(2009)]{Kunert09}
Kunert-Bajraszewska, M. \& Thomasson, P.: 2009, AN, 330, 210

\bibitem[Ku\-nert-\-Baj\-ra\-szew\-ska et al.(2010)]{Kunert10a}
Kunert-Bajraszewska, M.,Gawro\'nski, M.~P., Labiano, A., Siemiginowska, A.:
2010, MNRAS, 408, 2261

\bibitem[Ku\-nert-Baj\-ra\-szew\-ska \& Labiano(2010)]{Kunert10b}
Kunert-Bajraszewska, M., Labiano, A.: 2010, MNRAS, 408, 2279

\bibitem{} Kunert-Bajraszewska, M., Janiuk, A.: 2011, ApJ, 736, 125

\bibitem[Ku\-nert-\-Baj\-ra\-szew\-ska et al.(2013)]{kun2013}
Kunert-Bajraszewska, M., Siemiginowska, A., Labiano, A.: 2013, ApJ, 772, L7

\bibitem[Ku\-nert-\-Baj\-ra\-szew\-ska et al.(2014)]{kun2014}
Kunert-Bajraszewska, M., Labiano, A., Siemiginowska, A., Guainazzi, M.: 2014, MNRAS, 437, 3063

\bibitem[Labiano (2008)]{Labiano08letter} Labiano, A.: 2008, A\&A, 488, 59

\bibitem{} Maccagni, F. M., Morganti, R., Oosterloo, T. A., Mahony, E. K.: 2014, A\&A, 571, 67

\bibitem{} Maciel, T., Aleksander, P.: 2014, MNRAS, 442, 3469

\bibitem[Ma\-rec\-ki et al.(2003)]{Marecki03} Marecki, A., Spencer, R.~E., \&
Kunert, M.: 2003, PASA 20, 46

\bibitem{} Morganti, R., Killeen, N. E. B., Tadhunter, C. N.: 1993, MNRAS, 263, 1023

\bibitem[O'Dea \& Baum(1997)]{O'Dea97} O'Dea, C.~P., \& Baum, S.~A.: 1997,
AJ 113, 148

\bibitem{} Perucho, M., Quilis, V., Mart\'i, J.M.: 2011, ApJ, 743, 4

\bibitem[Readhead et al.(1996)]{Readhead96a} Readhead, A.~C.~S., Taylor, G.~B., Xu,
W., et~al.: 1996, ApJ 460, 612

\bibitem[Reynolds \& Begelman(1997)]{rb97} Reynolds, C. S., \& Begelman,
         M. C.: 1997, ApJ, 487, L135
         
\bibitem{} Sadler, E. M., Ekers, R. D., Mahony, E. K., Mauch, T., Murphy, T.: 2014, MNRAS, 438, 796 

\bibitem[Saikia et~al.(2001)]{saikia01} Saikia, D.~J., Jeyakumar, S.,
Salter, C.~J., et~al.: 2001, MNRAS 321, 37

\bibitem{} Schawinski, K., Koss, M., Berney, S., Sartori, L.: 2015, MNRAS, 451, 2517

\bibitem{} Shulevski, A., Morganti, R., Barthel, P.D. et al.: 2015, A\&A, 579, 27

\bibitem{} Siemiginowska, A., LaMassa, S., Aldcroft, T. L., Bechtold, Jill; Elvis, M.: 2008, ApJ, 684, 811
         
\bibitem[Snellen et al.(2000)]{Snellen00} Snellen, I.~A.~G., Schilizzi, R.~T.,
Miley, G.~K., de Bruyn, A.~G., Bremer, M.~N., R\"ottgering, H.~J.~A.: 2000,
MNRAS, 319, 445

\bibitem{} Tengstrand, O., Guainazzi,
M., Siemiginowska, A., Fonseca Bonilla, N., Labiano, A., Worrall, D. M.,
Grandi, P., Piconcelli, E.: 2009, A\&A, 501, 89

\bibitem{} Turner, R. J., Shabala, S. S.: 2015, ApJ, 806, 59

\bibitem{} Wagner, A. Y., Bicknell, G. V., Umemura, M.: 2012, ApJ, 757, 136

\bibitem[White et~al.(1997)]{white97} White, R.~L., Becker, R.~H., Helfand,
D.~J., \& Gregg, M.~D.: 1997, ApJ 475, 479


\end{thebibliography}
\end{document}